\documentclass[12pt]{article}  

\begin{document}

\newcounter{rown}
\def\bl{\setcounter{rown}{\value{equation}}
        \stepcounter{rown}\setcounter{equation}0
        \def\theequation{\thesection.\arabic{rown}\alph{equation}}
        }
\def\el{\setcounter{equation}{\value{rown}}
        \def\theequation{\thesection.\arabic{equation}}
        }
\def\sec{\setcounter{equation}0}
\renewcommand{\theequation}{\thesection.\arabic{equation}}

\title{Two-parameter extensions of the $\kappa$-Poincar\'{e}
 quantum  deformation }

\author{J. Lukierski$^\dag$ and V.D. Lyakhovsky$^\ddag$
\\ \\
 $^\dag$Institute for Theoretical Physics \\ University of Wroc\l{}aw \\
 pl. M. Borna 9, 50-204 Wroc{\l}aw, Poland
 \\
 e-mail:lukier@ift.uni.wroc.pl
\\ 
\\
$^\ddag$Departament of Theoretical Physics \\
 Sankt-Petersburg University  \\
 Ulianovskaya 1, Petrodvoretz, \\
 198904 Sankt Petersburg, Russia \\
 e-mail:lyakhovs@pobox.spbu.ru }

\date{}
\maketitle


\begin{abstract}
We consider the extensions of classical $r$-matrix for
$\kappa$-deformed Poincar\'{e} algebra which satisfy
 modified Yang-Baxter equation.
Two examples introducing additional deformation parameter
 (dimensionfull  $\frac{1}{\widetilde{\kappa}}$ or
dimensionless $\xi$) are presented.
We describe the
corresponding quantization (two-parameter $\kappa$-Poincar\'{e}
quantum  Hopf algebras) in explicite form as obtained by
twisting of  standard
$\kappa$-deformed framework.
 In the second example
quantum twist function depends on nonclassical generators, with
 $\kappa$-deformed coproduct. Finally we mention also the ``soft''
 twists with carrier in fourmomenta sector.
\end{abstract}

\section{Introduction}
Recently many authors (see e.g. \cite{lulyr1}--\cite{lulyr13})
considered the deformations of relativistic
  symmetries in the Hopf-algebraic
framework of quantum groups \cite{lulyr14}--\cite{lulyr16}.
 In particular physically
 appealing  are the $\kappa$-deformations
\cite{lulyr1}, \cite{lulyr3,lulyr4}, \cite{lulyr8,lulyr9},
\cite{lulyr13}
which introduce mass-like
deformation parameter $\kappa$ usually linked with Planck
mass $M_P$ (e.g. $\kappa=M_P$)  as well as with the quantum gravity
corrections.\footnote{The kinematic consequences of such
deformation has been studied in the framework of so-called doubly
special relativity (DSR) theories
 (see. e.g. \cite{lulyr17}--\cite{lulyr19}).} It
appears interesting to look for  multiparameter extensions of
standard $\kappa$-deformed framework and present the result as
multiparameter  Hopf-algebraic structure.

Quantum groups infinitesimally are characterized
 by the  classical
$r$-matrices. In particular standard $\kappa$-deformation of
Poincar\'{e} algebra \cite{lulyr1}, \cite{lulyr3,lulyr4},
\cite{lulyr8,lulyr9}
 is described by the following classical
$r$-matrix ($i=1,2,3$)

\begin{equation}\label{luly1}
r = \frac{1}{\kappa} \, N_i \wedge  P_i \, ,
\end{equation}
where $M_{\mu\nu} = (M_i = \frac{1}{2}\epsilon_{ijk}\, M_{jk}$,
$N_i = M_{i0})$ denote the Lorentz algebra generators and $P_\mu = (P_i,
P_0)$ represent the Abelian  fourmomenta. More
explicitly
\bl
\begin{eqnarray}\label{luly2a}
\left[ M_i , M_j \right]  & = & i \, \epsilon_{ijk} \, M_k  \, ,
\qquad\left[ N_i , N_j \right]   =  - i \, \epsilon_{ijk} \, M_k  \, ,
\\[5pt]
 & \phantom{=} & \left[ M_i , N_j \right]
 =  i \,\epsilon_{ijk} \, N_k  \, ,
\nonumber \\[8pt]
\left[ M_i , P_j \right]  & = & i \, \epsilon_{ijk} \, P_k  \, ,
\qquad  [ M_i , P_0 ] = 0 \, ,
\label {luly2b}
\\[5pt]
\left[ N_i , P_j \right]  & = & i \, \delta_{ij} \, P_0  \, ,
\ \qquad [N_i , P_0 ] = i P_i \, .
\nonumber
\end{eqnarray}
\el

The classical $r$-matrix (\ref{luly1}) satisfies the following
modified Yang-Baxter equation (MYBE)

\begin{equation}\label{luly3}
    \left[ \left[ r , r \right] \right] \equiv
    \left[ r_{12} , r_{13} \right] + \ldots
    = \frac{1}{\kappa^{2}} \, M_{\mu\nu} \wedge
    P^{\mu} \wedge P^{\nu} \, ,
\end{equation}
where $r_{12} = \frac{1}{\kappa}N_i \wedge P_i \wedge 1$,
 $r_{13} = \frac{1}{\kappa}N_i \wedge 1 \wedge P_i$ and
 $r_{23}= \frac{1}{\kappa} 1 \wedge N_i \wedge P_i$.
We found that the following ansatz
 provides new solutions of (\ref{luly3}) ($\xi_1$ and $\xi_2$ have the iniverse mass dimensions)
 \begin{equation}\label{luly4}
    \widetilde{r}= r + \delta \,  r \qquad
\delta \,  r  = \xi_1 \,
M_3 \wedge \, P_0 + \xi_2 M_3 \wedge P_+ \, ,
\end{equation}
where $P_{\pm} = P_1 \pm i P_2$. The relation

\begin{equation}\label{luly5}
    \left[ \left[ r ,\delta r \right] \right]+
 \left[ \left[\delta r , r \right] \right] +
  \left[ \left[\delta r ,\delta r \right] \right] = 0\, ,
    \end{equation}
is valid if
\begin{equation}\label{luly4bis}
    \xi_2 (\xi_1 - \frac{1}{\kappa})= 0 \, ,
\end{equation}
and the classical $r$-matrices
    satisfying  again the MYBE given by (\ref{luly3})
 are the following
\bl

\begin{eqnarray}\label{luly7a}
{\rm i)}  \quad
 \xi_2 = 0, \quad \xi_1=\frac{1}{\widetilde{\kappa}}\quad \quad
\phantom{ \hbox{($\xi$ is dimensionless)}}
&&
\cr\cr
 \delta r_1 =\frac{1}{\widetilde{\kappa}} M_3 \wedge P_0\, ,
\qquad \qquad \quad
&&
\end{eqnarray}
This modification of $\kappa$-Poincar\'{e}
 bi-algebra has been also mentioned in
\cite{lulyr20}.

\begin{eqnarray}\label{luly7b}
{\rm ii)} \quad
 \xi_2 = \xi \, \frac{1}{\kappa}, \quad
 \xi_1 =  \frac{1}{\kappa}
\quad \hbox{($\xi$ is dimensionless)} &&
\cr\cr
\qquad \qquad \delta r_2 = \frac{1}{\kappa} (M_3 \wedge P_0 + \xi
M_3 \wedge P_+)\,. &&
\end{eqnarray}
\el

    The aim of our paper is to describe explicitly  the quantization of the
    classical $r$-matrices described by
 (\ref{luly4}) and (\ref{luly7a}--\ref{luly7b}) by providing corresponding
  Hopf algebra formulae.

We shall introduce the Drinfeld twist function or twisting
element ${\mathcal{F} } $
which satisfies the $\kappa$-deformed twist equations \cite{lulyr21}
\bl
\begin{equation}\label{luly6a}
    {\mathcal{F}}_{12}(\Delta_{\kappa} \otimes {\rm id})
{\mathcal{F} } = {\mathcal{F} }_{23} ({\rm id} \otimes \Delta_{\kappa})
{\mathcal{F} }\, ,
\end{equation}
with the following linear term in the series expansion in deformation
 parameters (${\delta \, r}=
 {\delta\, r}_{k}^{(1)} \wedge {\delta \, r}^{(2)}_{k}$)
\begin{equation}\label{luly6b}
    {\mathcal{F} }= 1 \otimes 1 +
{\delta \, r}^{(1)}_k \otimes {\delta \, r}^{(2)}_k  + \ldots \ .
\end{equation}
\el
Twisting element ${\mathcal{F} } \in U_{\kappa} ({\mathcal{P}  }^4) \otimes U_{\kappa}
({\mathcal{P}  }^4) $ maps the $\kappa$-deformed Poincar\'{e}-Hopf algebra
 ${\mathcal{A}  }_{\kappa}(m,\Delta_{\kappa},S_{\kappa},\eta, \varepsilon)$
into two-parameter $\kappa$-deformed Poincar\'{e}-Hopf algebra
 ${\mathcal{A}  }_{\kappa,\alpha}(m,\Delta_{\kappa,\alpha},
 S_{\kappa,\alpha},\eta, \varepsilon)$
 where $ \alpha = \frac{1}{\widetilde{\kappa}}$ or $ \alpha =\xi$
\begin{equation}\label{luly7}
    \Delta_{\kappa, \alpha} = {\mathcal{F} }
\Delta_{\kappa}(a)
{\mathcal{F} }^{-1}\, ,
\end{equation}
\begin{equation}\label{luly8}
    S_{\kappa,\alpha}(a) = u \, S_{\kappa}(a)\, u^{-1}  \, ,
\end{equation}
 and
 $({\mathcal{F} }\equiv \sum\limits_{i} {\mathcal{F} }^{(1)}_i \otimes
 {\mathcal{F} }^{(2)}_i)$.
\begin{equation}\label{luly1.11}
u=\sum\limits_{i} {\mathcal{F} }^{(1)}_i \cdot S({\mathcal{F} }^{(2)}_i)\, .
\end{equation}
It appears that our twist function ${\mathcal{F} }$ satisfies more
specific factorized twist equations~\cite{lulyr22}
\bl
\begin{eqnarray}\label{luly9a}
\left(
\Delta_{\kappa} \otimes {\rm id}
\right)
{\mathcal{F} } & = & {\mathcal{F} }_{13} \cdot {\mathcal{F} }_{23} \, ,
\\[8pt]
\label{luly9b}
\left(
{\rm id} \otimes \Delta_{\kappa, \alpha}
\right)  {\mathcal{F} }
& = &
{\mathcal{F} }_{12} \, \cdot  {\mathcal{F} }_{13}\, ,
\end{eqnarray}
\el
\noindent
where in eq. (\ref{luly9b}) one should insert the
twisted coproduct (\ref{luly7}).
Indeed,
 one can show for the general case of deformed coproduct
(\ref{luly7}) that the relation
(\ref{luly6a}) follows from (\ref{luly9a}--\ref{luly9b}).

The plan of our paper is the following:

In Sect.~2 we describe the standard $\kappa $-deformation using the
bicrossproduct basis and find the solutions to the twist equations
corresponding to the terms
$\delta r_{1}$ (see \ref{luly7a})
and $\delta r_{2}$ (see \ref{luly7b}).
In Sect.~3 the
two-parameter $\kappa $-deformed Poincar\'{e} algebra is constructed in the
explicit form. In Sect. 4 we comment on the possibility of inserting
 ``soft" twist, with carrier algebra described by a fourmomentum sector.

We would like to stress here that usually the twists are classical, i.e.
 the carrier algebra of the twist function is a classical
Lie algebra, with primitive coproduct. Recently however, there were also
considered twists of quantum algebras, e.g.
 quantum Jordanian twist of the $q$-deformed $sl(2)$
 Borel subalgebra \cite{lulyr23};
 see also \cite{lulyr24}--\cite{lulyr25}.
 In this paper by quantizing classical
 $r$-matrix (\ref{luly7b}) we provide another example of quantum
 (i.e. nonclassical) twist.

\section{$\kappa$-deformed Poincar\'{e} algebra in
bicrossproduct basis and the twist function ${\mathcal{F} }$}
\setcounter{equation}{0}

The $\kappa$-deformed Poincar\'{e} Hopf algebra $A_\kappa$
 ($m, \Delta_\kappa, S_\kappa, \eta, \varepsilon$) can
 be written in different bases e.g. the standard one \cite{lulyr4},
bicrossproduct basis \cite{lulyr9} and the classical Poincar\'{e} algebra
 basis \cite{lulyr26}. Usually the twist is performed in classical Lie
 algebra basis; here we propose to consider the twist in
 nonclassical bicrossproduct basis.

The $\kappa$-deformed bicrossproduct basis is described in algebraic
sector by

i) classical Lorentz algebra (see (\ref{luly2a}))

ii) the cross relations of Lorentz generators and fourmomenta
 (\ref{luly2b}) with only one deformed relation

\begin{equation}\label{luly2.1}
    [N_i, P_j ] = i \, \delta_{ij}
\left[
\frac{\kappa}{2}
\left( 1 - e^{-\frac{2P_0}{\kappa}} \right)
+ \frac{1}{2\kappa}\, \overrightarrow{P}
\right]
- \frac{i}{\kappa} \, P_i \, P_j \, .
\end{equation}
The coalgebra sector is described by the following coproducts:
\begin{eqnarray}\label{luly2.2}
\Delta(P_0) & = & P_0 \otimes 1 + 1 \otimes P_0 \, ,
\nonumber\\[3pt]
\Delta(P_i) & = & P_i \otimes e^{- \frac{P_0}{\kappa} }
 + 1 \otimes P_i \, ,
\nonumber\\[3pt]
\Delta(M_i) & = & M_i \otimes 1 + 1 \otimes M_i \, ,
\nonumber\\[3pt]
\Delta(N_i) & = & N_i  \otimes
 e^{- \frac{P_0}{\kappa} } + 1 \otimes N_i -
\frac{1}{\kappa} \varepsilon_{ijk}
 M_{j} \otimes P_{k} \, .
\end{eqnarray}
The solution of the twist equations (\ref{luly6a}--\ref{luly6b})
 has the following form:

a) For the classical $r$-matrix (\ref{luly7a})

\begin{equation}\label{luly2.3a}
{\mathcal{F} }_{1}(\widetilde{\kappa}) =
 e^{\frac{1}{\widetilde{\kappa}}M_3 \otimes {\mathcal{P}  }_0}\, .
\end{equation}

Using (\ref{luly1.11}) one gets
\begin{equation}\label{luly2.4bis}
u(\widetilde{\kappa}) = e^{- \frac{1}{\widetilde{\kappa}} M_3\, P_0}
\end{equation}

b) For the classical $r$-matrix (\ref{luly7b})

\begin{eqnarray}\label{luly2.3b}
 {\mathcal{F} }_{2}(\kappa, \xi) & = &
 e^{\frac{1}{\kappa}M_3 \otimes P_0}
e^{M_3 \otimes \ln (1+ \xi P_+)}
\cr
& = &                                                                     
e^{M_{3}\otimes \ln \left(                                                
 e^{\frac{P_{0}}{\kappa }} + \xi\,  P_{+}\, e^{\frac{P_{0}}{\kappa}}      
 \right) } \, ,                                                           
\end{eqnarray}
and further

\begin{equation}\label{luly2.6bis}
 u(\kappa ,\alpha )=e^{M_{3}\ln \left( e^{-\frac{1}{\kappa }P_0}-\xi
P_{+}\right) }\, .
\end{equation}

The twist (\ref{luly2.3a}) satisfies (\ref{luly9a}--b)
 as follows from the general property of
commuting primitive generators \cite{lulyr22}. The solution (\ref{luly2.3b})
is less trivial. To check its validity consider the $3$-dimensional Hopf
algebra $\mathcal{A}=(C,D,H)$

\begin{equation}
\lbrack H,D]=D\,,\qquad \lbrack C,H]=[C,D]=0\,,  \label{luly2.5}
\end{equation}
with nonprimitive coproducts
\begin{eqnarray}\label{luly2.8bbis}
& \Delta (H)=H\otimes 1+1\otimes H\,,
 \qquad \Delta (C)=C\otimes C\,, &
\nonumber \\[5pt]
& \Delta
(D)=D\otimes 1+C\otimes D\,.  & \label{luly2.6}
\end{eqnarray}
We assert that the function
\begin{equation}
\mathcal{F}=e^{H\otimes \ln (C+D)}\,.  \label{luly2.7}
\end{equation}
is the twisting element for $\mathcal{A}$ and satisfies the factorised
equations (\ref{luly9a}--b). In fact the first of these equations is satisfied
due to primitivity of $H$.
To verify the second equation it is sufficient
to prove that the twisted coproduct
$\mathcal{F}\left( \Delta (C+D)\right)
\mathcal{F}^{-1}$ is group-like. Such a statement
  is the consequence of the relations (2.7--2.8).

In our case the Hopf algebra $\mathcal{A}=(\left\{ C,D,H\right\}
,\Delta =\Delta _{\kappa })$ is represented by

\begin{equation}
C=e^{\frac{P_{0}}{\kappa }}\,,\qquad D=P_{+}e^{\frac{P_{0}}{\kappa }
}\,,\qquad H=M_{3}\,.  \label{luly2.4}
\end{equation}
Thus we have demonstrated that $\mathcal{F}_{2}$ is the twisting element for
the $\kappa $-deformed Poincar\'{e} algebra.

\section{Two-parameter $\protect\kappa$-deformed Poincar\'{e}
algebra}

\setcounter{equation}{0}

{\bf i) Reshetikhin twist (\ref{luly2.3a})
(two mass-like deformation parameters
  $\kappa$ and $\widetilde{\kappa}$):}

In such a case the modified coproducts
 $\Delta _{\kappa ,\widetilde{\kappa }}
$ look as follows (we introduce $M_{\pm }=M_{1}\pm i\,M_{2}$ , $N_{\pm
}=N_{1}\pm i\,N_{2}$ and $P_{\pm }=P_{1}\pm i\,P_{2}$ )

\begin{equation}
{\begin{array}{lll}
\Delta _{\kappa ,\widetilde{\kappa }}\left( M_{\pm }\right)  & = & M_{\pm
}\otimes e^{\pm \frac{1}{\widetilde{\kappa }}P_{0}}+1\otimes M_{\pm },
\\[8pt]
\Delta _{\kappa ,\widetilde{\kappa }}\left( M_{3}\right)  & = & M_{3}\otimes
1+1\otimes M_{3},
 \\[8pt]
\Delta _{\kappa ,\widetilde{\kappa }}\left( N_{\pm }\right)  & = & N_{\pm
}\otimes e^{\left( \pm
 \frac{1}{\widetilde{\kappa }}-\frac{1}{\kappa }
\right) P_{0}}+1\otimes N_{\pm }+i\left( \mp
{  \frac{1}{\widetilde{\kappa }} } -
{ \frac{1}{\kappa }}\right) M_{3}\otimes P_{\pm }
\\[8pt]
&  & \pm i
{\displaystyle \frac{1}{\kappa }}
M_{\pm }\otimes P_{3}e^{\pm \frac{1}{\widetilde{\kappa }}P_{0}},
\\[8pt]
\Delta _{\kappa ,\widetilde{\kappa }}
\left( N_{3}\right)  & = & N_{3}\otimes
e^{-\frac{1}{\kappa }P_{0}}+1\otimes N_{3}-i
{\displaystyle \frac{1}{\widetilde{\kappa }}}%
M_{3}\otimes P_{3}
\\[8pt]
&&
-i {\displaystyle \frac{1}{2\kappa }}\, M_{+}\otimes P_{-}e^{\frac{1}{%
\widetilde{\kappa }}P_{0}}
+i {\displaystyle \frac{1}{2\kappa }}\,
 M_{-}\otimes P_{+}e^{-\frac{1}{\widetilde{\kappa }}%
P_{0}},
\\[8pt]
\Delta _{\kappa ,\widetilde{\kappa }}\left( P_{\pm }\right)  & = & P_{\pm
}\otimes e^{\left( \pm \frac{1}{\widetilde{\kappa }}-\frac{1}{\kappa }%
\right) P_{0}}+1\otimes P_{\pm },
\\[8pt]
\Delta _{\kappa ,\widetilde{\kappa }}\left( P_{3}\right)  & = & P_{3}\otimes
e^{-\frac{1}{k}P_{0}}+1\otimes P_{3},
\\[8pt]
\Delta _{\kappa ,\widetilde{\kappa }}\left( P_{0}\right)  & = & P_{0}\otimes
1+1\otimes P_{0}.
\end{array}}
\label{luly3.1}
\end{equation}

Using (\ref{luly2.4bis}) one gets the following formulae for
 antypodes:
\begin{equation}
\begin{array}{lll}
S_{\kappa ,\widetilde{\kappa }}\left( M_{\pm }\right)  & = & -M_{\pm }e^{\mp
\frac{1}{\widetilde{\kappa }}P_{0}},
 \qquad \qquad\qquad
S_{\kappa ,\widetilde{\kappa }}\left( M_{3}\right)   =  -M_{3},
\\[8pt]
S_{\kappa ,\widetilde{\kappa }}\left( N_{\pm }\right)  & = & -\left( N_{\pm
}-i\left( \mp \frac{1}{\widetilde{\kappa }}-\frac{1}{\kappa }\right)
M_{3}P_{\pm }\mp i\frac{1}{\kappa }M_{\pm }P_{3}\right) e^{\left( \mp \frac{1%
}{\widetilde{\kappa }}+\frac{1}{\kappa }\right) P_{0}},
\\[8pt]
S_{\kappa ,\widetilde{\kappa }}\left( N_{3}\right)  & = & -\left( N_{3}+i%
\frac{1}{\widetilde{\kappa }}M_{3}P_{3}+i\frac{1}{2\kappa }\left(
M_{+}P_{-}-M_{-}P_{+}\right) \right) e^{\frac{1}{\kappa }P_{0}},
\\[8pt]
S_{\kappa ,\widetilde{\kappa }}\left( P_{\pm }\right)  & = & -P_{\pm
}e^{\left( \mp \frac{1}{\widetilde{\kappa }}+\frac{1}{\kappa }\right) P_{0}},
\\[8pt]
S_{\kappa ,\widetilde{\kappa }}\left( P_{3}\right)  & = & -P_{3}e^{\frac{1}{k%
}P_{0}},
\qquad \qquad \qquad \qquad
S_{\kappa ,\widetilde{\kappa }}\left( P_{0}\right)  =  -P_{0}.
\end{array}
\label{anti}
\end{equation}

Only two generators $(P_{0},M_{3})$ have primitive coproducts  . It
should be noticed that if $\kappa =\widetilde{\kappa }$ $\left( \kappa =-%
\widetilde{\kappa }\right) $ then additionally the generator $P_{+}$ $\left(
P_{-}\right) $ becomes primitive.

{\bf ii) Complex twist (\ref{luly2.3b}) (one mass-like deformation parameter $%
\kappa$ and one dimensionless $\xi$).}

One gets the following coproducts ($\omega(\xi) =
\ln (1+ \xi P_+ )$)

\begin{equation}\label{luly3.3}
{\begin{array}{lll}
\Delta_{\kappa, \xi}(M_{+}) & = &
M_{+} \otimes e^{\frac{1}{\kappa}P_0 +\omega(\xi)} + 1 \otimes
 M_{+} \, ,
 \\[8pt]
\Delta_{\kappa, \xi} (M_{-}) & = &
M_{-} \otimes e^{-\frac{1}{\kappa}P_0 -\omega(\xi)} + 1 \otimes
 M_{-} + 2 \xi M_{3} \otimes P_{3} \, e^{-\omega(\xi)} \, ,
  \\[8pt]
\Delta_{\kappa, \xi} (M_{3}) & = &
M_{3} \otimes e^{-\omega(\xi)} + 1 \otimes
 M_{3} \, ,
 \\[8pt]
\Delta _{\kappa ,\xi }(N_{+}) &= &
N_{+}\otimes e^{\omega \left( \xi \right)
}+1\otimes N_{+}{-\frac{i}{\kappa }}\,M_{3}\otimes P_{+}\left( 1+e^{\omega
\left( \xi \right) }\right)
 \\[8pt]
&&
 + {\frac{i}{\kappa }}\,M_{+}\otimes P_{3}e^{\frac{%
P_{0}}{\kappa }+\omega \left( \xi \right) },
 \\[8pt]
\Delta_{\kappa, \xi} (N_{-}) & = &
N_{-} \otimes e^{-\frac{2P_0}{\kappa}-\omega(\xi)}
+ 1 \otimes N_{-} -
i \xi M_{3} \otimes \Big(
2 {\mathcal{P}  }_{0} e^{-\omega(\xi)} + {\displaystyle  \frac{P_{-}}{\kappa} }\,
 P_{+}
 \Big)
 \\[8pt]
&&
{\displaystyle - \frac{i}{\kappa}}\,  M_{-} \otimes
 P_{3}\,  e^{-\frac{P_0}{\kappa}- \omega(\xi)}\, ,
 \\[8pt]
\Delta_{\kappa, \xi} (N_{3}) & = &
N_{3} \otimes e^{-\frac{P_0}{\kappa}}
+ 1 \otimes N_{3}
 - {\displaystyle  \frac{i}{\kappa}}\, M_3 \otimes P_{3} e^{-\omega(\xi)}
 \\[8pt]
&&
+{\displaystyle  \frac{i}{2\kappa} }
\Big(
M_{-} \otimes P_{+} e^{-\frac{P_0}{\kappa}-\omega(\xi)} -
M_{+} \otimes P_{-}
e^{\frac{P_0}{\kappa}+\omega(\xi)}
\Big)
\, ,
 \\[8pt]
\Delta_{\kappa, \xi} (P_{+}) & = &
P_{+} \otimes e^{\omega(\xi)} + 1 \otimes P_{+} \, ,
 \\[8pt]
\Delta_{\kappa, \xi} (P_{-}) & = &
P_{-} \otimes e^{-\frac{2P_0}{\kappa}- \omega}
 + 1 \otimes P_{-} \, ,
 \\[8pt]
\Delta_{\kappa, \xi} (P_{3}) & = &
P_{3} \otimes e^{-\frac{2P_0}{\kappa}}
 + 1 \otimes P_{3} \, ,
 \\[8pt]
\Delta_{\kappa, \xi} (P_{0}) & = &
P_0 \otimes 1 + 1 \otimes P_0 \, .
\end{array} }
\end{equation}
Using (\ref{luly2.6bis}) and (\ref{luly8}) one can
 calculate also the antipodes.

We see that only one coproduct of the Poincar\'{e} generators,
  for $P_0$, remains primitive. Second primitive product is provided
   by the function $\omega(\xi)$ belonging to the enveloping algebra of
   ${\mathcal{P}  }_4$.

It should be added that the twist (\ref{luly2.3b}) is not unitary,
 i.e. it extends the real coproducts (\ref{luly2.2}) in a way
 which does not preserve the reality of the Poincar\'{e} generators.

\section{Final Remarks}
\setcounter{equation}{0}

The aim of this paper was to provide two examples of twisting
of standard $\kappa$-Poincar\'{e} algebra. In accordance with twisting
 quantization scheme only the coproducts and atipodes are
modified. The second twist is less attractive for physical
 applications (the coproducts become complex), but provides an interesting
 example of a twist with deformed carrier algebra.

In this note we neglected the Abelian twists with carriers in the
translation generators sector. It is possible to perform such deformations in the
initial $\kappa $-Poincar\'{e} algebra by adding to (\ref{luly1}) the following
classical $r$-matrix ($\xi_i$ - constant three-vector):
\begin{equation}\label{luly4.1}
\delta \, \widetilde{r} = \frac{1}{\kappa} \xi_i\, P_0 \wedge P_i \, ,
\end{equation}
which can be achieved by twisting the $\kappa$-Poincar'{e} algebra by
 the following twist factor
\begin{equation}\label{luly4.2}
 \mathcal{F}_{R}\left( \zeta _{i}\right)
=e^{\zeta _{i}P_{i}\otimes \left( e^{-\frac{1}{\kappa }P_{0}}-1\right) }\, .
\end{equation}
In the light of recent results \cite{lulyr27,lulyr28} which link such
type of twist with Seiberg-Witten $\theta$-deformation corresponding to
constant value of $\theta_{\mu\nu}$ in the commutator
 of space-time coordinates
($[\widehat{x}_{\mu},\widehat{x}_{\nu}]= i\theta_{\mu\nu}$),
such ``soft'' quantum deformation of $\kappa$-Poincar\'{e} algebra
might be also significant.


\subsection*{Acknowledgments}
One of the authors (VDL) is supported by the Russian Foundation
for Fundamental Research, grant N 03-01-00837.

\end{document}